\documentclass[preprint]{aastex}
\usepackage{epsf}

\def\be{\begin{equation}}
\def\ee{\end{equation}}
\def\ba{\begin{eqnarray}}
\def\ea{\end{eqnarray}}
\def\go{\mathrel{\raise.3ex\hbox{$>$}\mkern-14mu
             \lower0.6ex\hbox{$\sim$}}}
\def\lo{\mathrel{\raise.3ex\hbox{$<$}\mkern-14mu
             \lower0.6ex\hbox{$\sim$}}}
\def\br{{\bf r}}
\def\bv{{\bf v}}

\begin{document}

\title{Global Nonradial Instabilities of Dynamically Collapsing Gas Spheres}
\author{Dong Lai~\footnote{E-mail: dong@spacenet.tn.cornell.edu;
Alfred P. Sloan Research Fellow}}
\affil{Center for Radiophysics and Space Research, Department of
Astronomy, Cornell University,
Ithaca, NY 14853}

\begin{abstract}
Self-similar solutions provide good descriptions for the gravitational
collapse of spherical clouds or stars when the gas obeys a polytropic 
equation of state, $p=K\rho^\gamma$ (with $\gamma\le 4/3$, and $\gamma=1$
corresponds to isothermal gas). We study the behaviors of nonradial
(nonspherical) perturbations in the similarity solutions of Larson, Penston 
and Yahil, which describe the evolution of the collapsing cloud prior to
core formation. Our global stability analysis reveals the existence
of unstable bar-modes ($l=2$) when $\gamma\le 1.09$. In particular,
for the collapse of isothermal spheres, which applies to the early
stages of star formation, the $l=2$ density perturbation relative to
the background, $\delta\rho({\bf r},t)/\rho(r,t)$, increases
as $(t_0-t)^{-0.352}\propto \rho_c(t)^{0.176}$, where $t_0$ denotes
the epoch of core formation, and $\rho_c(t)$ is the cloud central density.
Thus, 
the isothermal cloud tends to evolve into an ellipsoidal shape 
(prolate bar or oblate disk, depending on initial conditions) as the 
collapse proceeds. This shape deformation may facilitate fragmentation of the
cloud. In the context of Type II supernovae, core collapse is described by
the $\gamma\simeq 1.3$ equation of state, and our analysis indicates  
that there is no growing mode (with density perturbation) 
in the collapsing core
before the proto-neutron star forms, although nonradial
perturbations can grow during the subsequent accretion
of the outer core and envelope onto the neutron star.

We also carry out a global stability analysis for the self-similar 
expansion-wave solution found by Shu, which describes the 
post-collapse accretion (``inside-out'' collapse) of isothermal gas
onto a protostar. We show that this solution is unstable to 
perturbations of all $l$'s, although the growth rates are unknown.  
\end{abstract}


\section{Introduction}

The gravitational collapse of molecular clouds leading to star formation
has long been an active area of study. In the early stages of collapse
(from $\rho\lo 10^{-19}$~g~cm$^{-3}$ to $\rho\sim 10^{-12}$~g~cm$^{-3}$)
the gas remains approximately isothermal (at temperature $\sim 10$~K) due to
efficient cooling by dust grains (see, e.g., Myhill \& Boss 1993).
The gas dynamics is then specified by two dimensional parameters, 
the gravitational constant $G$ and the isothermal sound speed $a$, so that the
flow is expected to approach a self-similar form in the asymptotic limit, 
when the memory of initial conditions is ``lost''. Larson (1969) and Penston
(1969) found a similarity solution which describes the pre-collapse (i.e.,
before the central protostar forms) evolution of the cloud, in which the gas
collapses from rest, accelerating until it cruises at Mach number of $3.3$ and
the density profile reaches a $r^{-2}$ power law. The Larson-Penston
solution contains a nonsingular homologous inner core and a
supersonic outer envelope. A qualitatively different set of similarity
solutions was found by Shu (1977). Of particular interest is Shu's
expansion-wave solution which describes
the post-collapse accretion of a singular isothermal gas cloud onto a
protostar. In this solution, the flow starts from hydrostatic equilibrium 
(with a $r^{-2}$ density profile) and a rarefaction wave expands
from the center and initiates the collapse (the so-called ``inside-out''
collapse); Inside the expansion-wave front, the flow eventually attains 
the free-fall behavior ($v\propto r^{-1/2}$) at small radii, with density
$\rho\propto r^{-3/2}$. The link between the Larson-Penston pre-collapse
solution and Shu's expansion-wave solution was elucidated by Hunter (1977), 
who showed that the Larson-Penston solution can be continued to the 
post-collapse phase and that there exists an infinite (but discrete) number 
of pre- and post-collapse solutions of a different type (called ``Type I'';
see \S 2), among which the expansion-wave solution represents 
a limiting case.  
Figure 1 illustrates the properties of different self-similar solutions 
for the collapse and accretion of isothermal spheres.\footnote{We note 
that Whitworth \& Summers (1985) have 
found a continuum of similarity solutions by relaxing the analyticity condition
of the flow at the transonic point; However, these generalized solutions are
locally unstable (Hunter 1986; Ori \& Piran 1988), and therefore may not be
realized in astrophysical situations. We also mention that Boily and
Lynden-Bell (1995) have constructed similarity solutions for the gravitational
collapse of radiatively cooling gas spheres (with emissivity having a 
power-law dependence on density and temperature).} 

With the plethora of possible similarity solutions, it is important 
to know which, if any, of them are actually realized by collapse 
of isothermal clouds. One-dimensional hydrodynamical simulations, 
starting from a regular (Bonner-Ebert) sphere, generally indicate that the
collapse resembles the Larson-Penston similarity form in the asymptotic
limit (Hunter 1977; Foster \& Chevalier 1993). This is consistent with
the recent finding of Hanawa \& Nakayama (1997), who showed that the
pre-collapse Type I solutions of Hunter's (see Fig.~1) are strongly unstable
against global spherical perturbations, and therefore are unlikely to be
realized in astrophysical situations or numerical simulations. 

Similarity solutions have also been investigated in the context of
core-collapse supernovae (Goldreich \& Weber 1980; Yahil 1983), 
where the equation of state of the collapsing iron core can be 
approximated by that of a polytrope, $p=K\rho^\gamma$, 
where $K$ is a constant and $\gamma\simeq 4/3$. 
(In fact, the effective $\gamma$
is about $1.3$ from the onset of electron capture to the neutrino trapping
density, i.e., for $4\times 10^9$~g~cm$^{-3}\lo\rho\lo 10^{12}$~g~cm$^{-3}$; 
$\gamma$ becomes close to $4/3$ when $\rho\go 10^{12}$~g~cm$^{-3}$ until
nuclear density is reached.) Goldreich \& Weber (1980) studied the
special case of $\gamma=4/3$, which provides a good description for the inner
homologous core; They also performed a global perturbation analysis
and showed that the inner core is stable against all radial and nonradial 
perturbations. Yahil (1983) generalized the Goldreich-Weber solution 
to general $\gamma\le 4/3$; this allows for a proper description of the outer
core which collapses supersonically. Since Yahil's solution is the
same as the Larson-Penston solution except for different values of
$\gamma$, we shall often refer them as Larson-Penston-Yahil solutions in the
remainder of this paper.

The similarity solutions described above (in both star formation and
supernova contexts) assume idealized spherical flows. A realistic gas cloud, 
however, contains nonradial (nonspherical) perturbations,\footnote{A realistic
flow may also contain a non-negligible amount of angular momentum and 
magnetic fields --- these are neglected in the main text of this paper. 
In Appendix A we discuss the perturbative effects of rotation on
Larson-Penston-Yahil solutions. Terebey,
Shu \& Cassen (1984) have considered how slow rotation affects the
expansion-wave solution, and Galli \& Shu (1993a,b) have studied the
perturbative effects of magnetic fields (see also Li \& Shu 1997).} 
and it is of interest to understand the behaviors of these perturbations 
during the collapse/accretion of the cloud.
In general, multi-dimensional hydrodynamical simulations are needed
to follow the evolution of the perturbed flow, especially when 
the perturbations become nonlinear. The large dynamical range involved in a
collapse makes such simulations particularly challenging (e.g., star 
formation ultimately involves collapse from $\rho\lo 10^{-19}$~g~cm$^{-3}$
to $\rho\go 0.1$~g~cm$^{-3}$; even the initial isothermal collapse stage
involves seven orders of magnitude increase in densities; 
see Truelove et al.~1997,1998 and Boss 1998 for a discussion on the numerical
subtleties). An alternative, complementary
approach is to carry out linear stability analysis to determine whether the
flow is unstable to the growth of any nonradial perturbations. 
Since the unperturbed flow varies in space and time in a self-similar manner, 
a global analysis is needed to study perturbations which vary on similar
scales as the unperturbed flow itself. The stability properties of the flow
therefore depend crucially on boundary conditions at different locations of
the flow. In this paper we perform global stability analysis for 
Larson-Penston-Yahil solutions (general $\gamma$) and for 
Shu's expansion-wave solution ($\gamma=1$ only) to determine 
whether these similarity flows contain growing nonradial modes. 

While the stability properties of isothermal similarity collapse solutions 
(the Larson-Penston solution and the expansion-wave solution) are 
relevant to the formation of binary (and multiple) stars (see \S 5), 
the present study stems from our attempts to understand the origin of
asymmetric supernovae and pulsar kicks (Goldreich, Lai \& Sahrling 1996;
see also Lai 1999). 
Numerical simulations indicate that local hydrodynamical instabilities 
in the collapsed stellar core (e.g., Burrows et al.~1995; Janka \& M\"uller
1994,~1996; Herant et al.~1994), which can in principle lead
to asymmetric matter ejection and/or asymmetric neutrino emission,
are not adequate to account for kick velocities $\go 100$~km~s$^{-1}$ 
(Burrows \& Hayes 1996; Janka 1998). Global asymmetric perturbations of
presupernova cores may be required to produce the observed kicks. 
Goldreich et al.~(1996) suggested that overstable g-modes 
driven by shell nuclear burning may provide seed perturbations 
which could be amplified during core collapse (see also Lai \& Goldreich 
2000). While the analysis of Goldreich \& Weber (1980) shows that the inner
homologous core is stable against nonradial perturbations, the situation is
not so clear for the supersonically collapsing outer core
where pressure plays a less important role. It is therefore
important to analyse the global stability of Yahil's self-similar solution. 
Hanawa \& Matsumoto (1999) have recently found a globally unstable 
bar mode in the pre-collapse Larson-Penston solution (for isothermal collapse).
Our independent calculations confirm their result for $\gamma=1$. 
Since the analysis of Hanawa \& Matsumoto is restricted to
perturbations with real eigenvalues and eigenfunctions (see \S 3), it is not
clear whether there exists any other growing modes, nor is it clear whether the
growing bar-mode persists for general values of $\gamma$ (see also 
Hanawa \& Matsumoto 2000a). 

The remainder of this paper is organized as follows. Section 2 summarizes
the basic properties of the (unperturbed) Larson-Penston-Yahil 
similarity solution. This serves as a preparation for our stability
analysis presented in Section 3. (For readers not interested in 
technical details, the main results are given in \S 3.4 and Figures 3-5.)
In Section 4 we show that Shu's expansion-wave solution for isothermal 
collapse is unstable to nonradial perturbations of all angular orders. 
Finally, we discuss the astrophysical implications of our results in \S 5. 
Appendix A contains a discussion of the rotational and vortex modes of 
Larson-Penston-Yahil solutions.

\section{Spherical Larson-Penston-Yahil Self-Similar Collapse}

Here we briefly review the (pre-collapse) Larson-Penston-Yahil similarity 
solutions for spherical collapse and summarize the basic flow properties
which are needed for our perturbation analysis (\S 3). 

We adopt a barotropic equation of state, where pressure $p$ and density
$\rho$ are related by $p=K\rho^\gamma$, and $K$ and $\gamma$ are constants. 
To have gravitational collapse we require $\gamma\le 4/3$. 
The two dimensional parameters of the problem are $K$ and
the Newton's constant $G$, from which we can construct a unique
similarity variable 
\be
\eta={r\over R(t)},\qquad
R(t)=K^{1/2}G^{(1-\gamma)/2}(-t)^{(2-\gamma)},
\ee
where $r$ is the spherical radius, and the time $t$ is measured from the epoch
of core formation (i.e., the center formally collapses to a singularity at
$t=0$). Our analysis in \S 3 will be restricted to the pre-collapse solutions,
so the domain of interest corresponds to $t<0$. 
From dimensional consideration, we can write the dynamical (dependent) 
variables of the flow in self-similar forms:
\ba
&\rho(r,t)=\rho_t D(\eta),\qquad & \rho_t=G^{-1}(-t)^{-2}\label{rhot}\\
&v(r,t)=v_tV(\eta),\qquad & v_t=K^{1/2}G^{(1-\gamma)/2}(-t)^{1-\gamma}\\
&p(r,t)=p_tP(\eta),\qquad & p_t=KG^{-\gamma}(-t)^{-2\gamma}\\
&\psi(r,t)=\psi_t\Psi(\eta),\qquad & \psi_t=KG^{1-\gamma}(-t)^{2(1-\gamma)}
\label{psit}\\
&m(r,t)=m_tM(\eta),\qquad & m_t=K^{3/2}G^{(1-3\gamma)/2}(-t)^{4-3\gamma}\\
&u(r,t)=u_tU(\eta),\qquad & u_t=KG^{1-\gamma}(-t)^{3-2\gamma}.\label{ut}
\ea
Here $v(r,t)$ is the radial velocity, $\psi(r,t)$ is the gravitational
potential and $m(r,t)$ is the mass interior to radius $r$; 
For later purpose, we have defined the velocity stream function
$u$ such that ${\bf v}=\nabla u$, and $V(\eta)=dU/d\eta=U'$;
here and hereafter we shall use prime ($'$)  to denote $d/d\eta$. 
In terms of the dimensionless variables, the equation of state 
is simply $P=D^\gamma$. The continuity equation,
Euler equation and Poisson equation become
\ba
&&WD'+DV'+2D\left(1+{V\over\eta}\right)=0,\label{cont}\\
&&\gamma D^{\gamma-2}D'+WV'+(\gamma-1)V+{M\over\eta^2}=0,\label{euler}\\
&&M'=4\pi\eta^2 D,\label{mass}
\ea
where we have used the relation $M=\eta^2\Psi'$ and have defined
\be
W\equiv V+(2-\gamma)\eta.
\ee
Note $v_tW=v-(dR/dt)\eta$ is simply the ``peculiar'' flow velocity with 
respect to the homologous frame. 
Equation (\ref{cont}) can be integrated out, with the help of equation
(\ref{mass}), to give
\be
4\pi\eta^2DW=(4-3\gamma)M.
\label{dwm}\ee
Eliminating $V'$ from equations (\ref{cont}) and (\ref{euler}), we obtain
\be
D'=D\left[(1-\gamma)W-{2W^2\over\eta}+{M\over\eta^2}
-(\gamma-1)(2-\gamma)\eta\right]\left(W^2-\gamma D^{\gamma-1}\right)^{-1}.
\label{dp}\ee
We see there is a sonic point at $\eta=\eta_s$, where
$W^2=\gamma D^{\gamma-1}$, i.e., the (dimensionless) peculiar velocity $W$
equals the sound speed $(\gamma D^{\gamma-1})^{1/2}$.

Equations (\ref{mass}), (\ref{dwm}) and (\ref{dp}) determine the spherical
self-similar flow. Some properties of the flow are as follows. 
For $\eta\rightarrow 0$:
\be
D\rightarrow D_0,\qquad
V\rightarrow -{2\over 3}\eta,\qquad
M\rightarrow {4\pi\over 3}D_0\eta^3;
\ee
For $\eta\rightarrow\infty$:
\be
D\propto \eta^{-2/(2-\gamma)},\qquad
V\propto \eta^{(1-\gamma)/(2-\gamma)},\qquad
V/A \rightarrow {\cal M}_\infty={\rm constant}.
\label{bcinfty}\ee
The physical solution is obtained by adjusting $D_0$ so that the flow passes
through the sonic point smoothly. To obtain accurate transonic solution, it is
useful to analyse the behavior of the flow near the sonic point. 
The values of $D_s=D(\eta_s),~W_s=W(\eta_s)$ and $M_s=M(\eta_s)$
are completely determined by requiring both the denominator and 
numerator of equation (\ref{dp}) to vanish at $\eta_s$. For 
$\epsilon=(\eta-\eta_s)/\eta_s\ll 1$, let
$D=D_s(1+\alpha \epsilon)$ and $W=W_s(1+\beta \epsilon)$.
From equations (\ref{mass}) and (\ref{dwm}), we find
$M=M_s[1+(2+\alpha+\beta)\epsilon]$ and
$2+\alpha+\beta=({\eta_s/W_s})(4-3\gamma)$. Taylor expansion of equation
(\ref{dp}) around $\eta_s$ then yields
\ba
&&(\gamma+1)\alpha^2-\left[(9-7\gamma){\eta_c\over W_s}-8\right]\alpha
+2-(\gamma-1)(2-\gamma){\eta_s^2\over W_s^2}\nonumber\\
&&\qquad +\left[(4-3\gamma){\eta_s\over W_s}-2\right]
\left[{M_s\over\eta_sW_s^2}+(1-\gamma){\eta_s\over W_s}-4\right]=0.
\label{root}\ea
For $\gamma=1$, the two roots are $\alpha=-1$ and $\alpha=\eta_s-3$, and the
former gives the Larson-Penston solution.
For general $\gamma$, the smaller of the two roots of (\ref{root})
corresponds the Yahil-Larsen-Penston solution, which is the only 
solution with $|V|$ supersonic at large $\eta$ (this is called type 
II solution by Hunter 1977). The other root gives rise to a infinite (but
discrete) number of solutions which are subsonic in $|V|$ at large $\eta$
(called Type I by Hunter). We will not discuss these type I solutions 
further in this paper since they are strongly unstable against radial 
perturbations (Hanawa \& Nakayama 1997). 
Our numerical procedure for finding the transonic solution is as
follows: Guess $D_0$ and $\eta_s$; 
Using the boundary conditions given above, integrate 
equations (\ref{mass}) and (\ref{dp}) outward from $\eta=0$ and 
inward from $\eta_s$ to a middle point $\eta_{\rm mid}$; 
Using the Newton-Raphson scheme (Press et al.~1992)
to vary $D_0$ and $\eta_s$ so that the two integrations match at 
$\eta_{\rm mid}$. 

Figure 2 gives two examples of the Larson-Penston-Yahil self-similar 
solutions of spherical collapse (for $\gamma=1$ and $1.3$). For convenience, 
we list in Table 1 the key parameters of the solutions
for different values of $\gamma$.

\section{Perturbations of Larson-Penston-Yahil Collapse Solution}

Our stability analysis relies on calculating the global linear modes
of the self-similar flow. In general, it is not meaningful to speak of 
modes in flows where the unperturbed state is time dependent. 
Self-similar flows constitute an exception, since the spatial structure of 
the unperturbed flow is constant in shape, although not in scale. 
In this case, a mode represents a linearized disturbance with shape-preserving 
spatial structure and power-law time dependence relative to the unperturbed
flow. The mode structure and stability depend on the feedback between
boundary conditions at different locations of the flow.  

\subsection{Perturbation Equations}

We consider flows with no net angular momentum and 
vorticity.\footnote{When rotation is a small perturbation, 
it is decoupled from the density perturbation and the potential flow. 
We discuss the rotational perturbations of self-similar flows in Appendix A.}
The fluid velocity is completely specified by the stream function
(velocity potential), i.e., ${\bf v}=\nabla u$. The continuity 
equation, Euler equation and Poisson equation for the
irrotational flow can be written as
\ba
&&{\partial\rho\over\partial t}+\nabla\cdot(\rho\nabla u)=0,\\
&&{\partial u\over\partial t}+{1\over 2}(\nabla u)^2+h+\psi=0,\\
&&\nabla^2\psi=4\pi G\rho,
\ea
where the enthalpy $h=\int\!dP/\rho=\gamma K\rho^{\gamma-1}/(\gamma-1)$
for $\gamma\neq 1$ and $h=K\ln\rho$ for $\gamma=1$.
The perturbed hydrodynamical equations are 
\ba
&&{\partial\over\partial t}\delta\rho+{1\over r^2}
{\partial\over\partial r}\left(r^2\delta\rho{\partial u\over\partial r}
\right)+{\partial\rho\over\partial r}{\partial\delta u\over\partial r}
+\rho\nabla^2\delta u =0,\label{pert1}\\
&&{\partial\over\partial t}\delta u+v{\partial \delta u\over\partial r}
+\gamma K\rho^{\gamma-2}\delta\rho+\delta\psi=0,\label{pert2}\\
&&\nabla^2\delta\psi-4\pi G\delta\rho=0,\label{pert3}
\ea
where $\delta\rho,~\delta u$ and $\delta\psi$ are the Eulerian 
perturbations of density, velocity potential and gravitational potential,
respectively. Separating out the angular dependence in terms of
spherical harmonics $Y_{lm}$, we write the perturbations in the form
\ba
&&\delta\rho(\br,t)=(-t)^s\rho_t\,\delta\!D(\eta)Y_{lm},\label{delrho}\\
&&\delta u(\br,t)=(-t)^s u_t\,\delta U(\eta)Y_{lm},\label{delu}\\
&&\delta \psi(\br,t)=(-t)^s\psi_t\,\delta \Psi(\eta)Y_{lm},\label{delpsi0}
\ea
where $\rho_t,~\psi_t$ and $u_t$ are given by equations (\ref{rhot}),
(\ref{psit}) and (\ref{ut}). The velocity perturbation is given by 
\be
\delta {\bf v}(\br,t)=\nabla\delta u(\br,t)=
(-t)^s v_t\left[\delta V_r(\eta)\,{\hat r}+\delta V_\perp(\eta)
\hat\nabla_\perp\right]\,Y_{lm},
\label{delv}\ee
where 
\be
\hat\nabla_\perp\equiv
\hat\theta{\partial\over\partial\theta}
+{\hat\phi\over\sin\theta}{\partial\over\partial\phi},
\ee
and the dimensionless radial and tangential velocity perturbations are
\be
\delta V_r(\eta)=\delta U'(\eta),\qquad
\delta V_\perp(\eta)={\delta U(\eta)\over\eta}.
\ee
In equations (\ref{delrho})-(\ref{delv}), the (unknown) power-law index $s$
constitutes an eigenvalue of the problem. 
Since $\delta\rho(\br,t)/\rho(r,t)=
(-t)^s\left[\delta\!D(\eta)/D(\eta)\right]Y_{lm}$ (and similarly for other
variables), the value of $s$ determines the global behavior of the
perturbation relative to the unperturbed flow: The perturbation is globally
unstable if the real part of $s$, Re($s$), is less than zero, and it is stable
if ${\rm Re}(s)>0$. Substituting (\ref{delrho})-(\ref{delpsi0})
into equations (\ref{pert1})-(\ref{pert3}), we have
\ba
&&W\delta D'+D\delta U''+\Bigl(2-s+V'+{2\over\eta}V\Bigr)\delta D
+\Bigl(D'+{2\over\eta}D\Bigr)\delta U'-{l(l+1)\over\eta^2}D\delta U=0,
\label{deld2}\\
&&W\delta U'+\gamma D^{\gamma-2}\delta D+(2\gamma-3-s)\delta U+\delta\Psi=0,
\label{delu}\\
&&\delta\Psi''={l(l+1)\over\eta^2}\delta\Psi-{2\over\eta}\delta\Psi'
+4\pi\delta D.\label{delpsi}
\ea
We can eliminate $\delta U''$ from (\ref{deld2}) using (\ref{delu}) to obtain
\ba
&&\left(W^2-\gamma D^{\gamma-1}\right){\delta D'\over D}
=-(3-3\gamma+s-2V')\delta U'\nonumber\\
&&\qquad -\left[\Bigl(2-s+V'+{2\over\eta}V\Bigr)WD^{-1}
-\gamma(\gamma-2)D^{\gamma-3}D'\right]\delta D\nonumber\\
&&\qquad +{l(l+1)\over\eta^2}W\delta U+\delta\Psi'.\label{deld}
\ea
Thus the perturbation equation is singular at the sonic point ($\eta=\eta_s$).
Equations (\ref{delu})-(\ref{deld}) are the basic
equations for the eigenvalue problem.

\subsection{Boundary Conditions}

To solve for the eigenvalue $s$, we need to know the boundary conditions. 
Since the unperturbed flow is regular at $\eta\rightarrow 0$, we require
the perturbations to be regular also. This gives, for $\eta\rightarrow 0$, 
\be
\delta D=\delta D_0\eta^l,\qquad
\delta U=\delta U_0\eta^l,\qquad
\delta\Psi=\delta\Psi_0\eta^l,\qquad\qquad\qquad (\eta\rightarrow 0)
\label{eta0}\ee
where the three constants $\delta D_0,~\delta U_0$ and $\delta\Psi_0$ are
related by
\be
\gamma D_0^{\gamma-2}\delta D_0=-\left[
\Bigl({4\over 3}-\gamma\Bigr)l-3+2\gamma-s\right]\delta U_0-\delta\Psi_0.
\label{eta00}\ee
Since the unperturbed flow is nearly static ($V\rightarrow 0$)
at $\eta\rightarrow 0$, the conditions (\ref{eta0})-(\ref{eta00}) 
are similar to those applied for nonradial pulsations in stars (e.g.,
Unno et al.~1989).  

The boundary conditions at $\eta\rightarrow\infty$ are trickier. 
Let $\delta D\propto \eta^a$, $\delta U\propto\eta^b$ and
$\delta \Psi\propto \eta^c$ for $\eta\rightarrow \infty$. 
There are four independent solutions to the
fourth order systems of differential equations. Using the scaling relations
in (\ref{bcinfty}), we find that the values of $a,~b,~c$ for the four 
solutions are
\ba
&{\rm Solution~I:}&\qquad
a={s-2\over 2-\gamma},\qquad
b=c=a+2={s+2(1-\gamma)\over 2-\gamma};\\
&{\rm Solution~II:}&\qquad
a={s-3\over 2-\gamma},\qquad
b={s+3-2\gamma\over 2-\gamma},\qquad
c=a+2={s+1-2\gamma\over 2-\gamma};\\
&{\rm Solution~III:}&\qquad
a=-l-3-{2\over 2-\gamma},\qquad
b=c=-(l+1);\\
&{\rm Solution~IV:}&\qquad
a=l-2-{2\over 2-\gamma},\qquad b=c=l.
\ea
For each solution, the ratio $\delta U/\delta\Psi$ and
$\delta D/\delta\Psi$ are uniquely determined. The general solution
of equations (\ref{delu})-(\ref{deld}) is a superposition of Solution
I-IV. In Solution I and II, the potential perturbation $\delta\Psi$ is 
produced by local density perturbation $\delta D$ (thus 
$c=a+2$); In Solution III, $\delta\Psi$ at a large
$\eta$ is produced by a multipole moment associated with $\delta D$ at smaller
$\eta$. Solutions I-III are physically allowed. Solution IV, however, 
is not arrowed, since it corresponds to the situation where $\delta \Psi$ at a
given (large) $\eta$ is produced by density perturbation at even larger $\eta$,
and $\delta\Psi$ increases without bound as $\eta\rightarrow\infty$.
Therefore, the eigenmode at larger $\eta$ is a linear combination 
of Solution I,II and III. Unless the Re$(s)$ is extremely negative, i.e.,
for ${\rm Re}(s)>-(4-3\gamma)-(2-\gamma)l$, 
the behaviors of $\delta D$ and $\delta\Psi$ at large $\eta$ are dominated by
Solution I, while the behavior of $\delta U$ is dominated by Solution II. Thus
we have 
\be
{\delta D\over D},~{\delta U\over U},~{\delta\Psi\over\Psi}
\propto \eta^{s/(2-\gamma)},
\label{deldd}\ee
where we have used 
$U\sim\eta V\propto \eta^{(3-2\gamma)/(2-\gamma)}$ and
$\Psi\sim \eta^2 D\propto \eta^{2(1-\gamma)/(2-\gamma)}$
at $\eta\rightarrow\infty$.
In practice, we implement the outer boundary condition at large $\eta$ as
\be
\delta\Psi'={c\over \eta}\delta\Psi,\qquad c={s+2(1-\gamma)\over 2-\gamma}
\qquad\qquad (\eta\rightarrow\infty).
\label{delpsip}\ee

Equation (\ref{deldd}) indicates that when ${\rm Re}(s)>0$, the fractional
perturbations $\delta D/D,~\delta U/U$ and $\delta\Psi/\Psi$ diverge
as $\eta$ increases to infinity. Thus only for globally unstable
modes (Re$(s)<0$) are the fractional perturbations finite at
$\eta\rightarrow\infty$. Whether such an unstable mode exists (for a given $l$
and $\gamma$) is unknown {\it a priori}. Note that equation (\ref{deldd}) also
corresponds to
\be
{\delta\rho(\br,t)\over\rho(r,t)},~
{\delta u(\br,t)\over u(r,t)},~
{\delta\psi(\br,t)\over\psi(r,t)}\propto r^{s/(2-\gamma)}(-t)^0,
\ee
i.e., the perturbations are independent of time for $\eta\rightarrow\infty$. 

Since the sonic point ($\eta=\eta_s$) is a singular point of equation
(\ref{deld}), another crucial condition for the perturbation analysis
is that the perturbations remain regular and pass through the sonic 
point smoothly.

\subsection{Numerical Method}

Our numerical procedure for finding an eigenmode is as follows:
(i) We first guess $s$ and $\delta U_0/\delta\Psi_0$ (note that 
in general they are complex), 
and use equation (\ref{eta00}) to find $\delta D_0/\delta\Psi_0$; 
(In plotting the eigenfunctions below, we adopt the normalization 
such that $\delta\Psi_0=1$);
(ii) We then integrate equations (\ref{delu})-(\ref{deld})
from a small $\eta_{\rm in}\ll 1$ to $\eta_s$ and then from $\eta_s$ to
a large $\eta_{\rm out}$ (we typically choose $\eta_{\rm out}=10^3-10^4$);
(iii) Using a Newton-Raphson scheme (Press et al.~1992), 
we vary the values of $s$ and 
$\delta U_0/\delta\Psi_0$ until the right-hand side of equation (\ref{deld})
vanishes at $\eta_s$ and condition (\ref{delpsip}) is satisfied at 
$\eta_{\rm out}$. Note that in step (ii), we first integrate the equations to
$\eta_{s-}=\eta_s(1-\varepsilon)$, where $0<\varepsilon\ll 1$ (we typically
choose $\varepsilon=10^{-4}-10^{-3}$), and advance the solution to 
$\eta_s$ and to $\eta_{s+}=\eta_s(1+\varepsilon)$ using the derivatives
evaluated at $\eta_{s-}$, and then continue the integration
from $\eta_{s+}$ to $\eta_{\rm out}$. We have found that this 
procedure works well except that for some high-order modes
the convergence of the eigenvalue $s$ as $\varepsilon$ decreases
requires very small $\varepsilon$. 
We have also tried using derivatives evaluated at $\eta_s$ (and using
L'H\^opital's rule to calculate $\delta D'$ at $\eta_s$) to advance the
solution from $\eta_{s-}$ to $\eta_{s+}$, but this did not lead to significant 
improvement. Ideally, one should not integrate ``into'' the singular point
$\eta_s$, but rather should integrate from $\eta_s$ inward to a
midpoint $\eta_{\rm mid}$ ($<\eta_s$) and match the solution there. 
However, this introduces several additional unknown parameters and makes
the multi-dimensional Newton-Raphson scheme difficult to converge 
in practice.

\subsection{Results}

We first note that for $l=1$, the lowest-order mode (the one
with no node in the radial eigenfunction) is a trivial 
solution; it corresponds to choosing the origin of the coordinates away from
the center of the spherical flow. The eigenfunctions are 
$\delta D=D',~\delta U=U',~\delta\Psi=\Psi'$. The negative eigenvalue
$s=\gamma-2$ should not be considered as an indication of global
instability. All other nonradial modes are nontrivial. 

\subsubsection{Unstable Modes}

For $\gamma=1$ and $l=2$, we find that the lowest-order mode 
has a real eigenvalue, $s=-0.352$.
Figure 3 depicts the eigenfunctions of the mode. Near the center
($\eta\rightarrow 0$), we find $\delta U_0/\delta\Psi_0=-0.836$. The
eigenfunctions are well-behaved everywhere, and go through the transonic point
$\eta_s$ smoothly. The negative eigenvalue $s$ indicates that the bar-mode is
globally unstable, with 
\ba
&&{\delta\rho(\br,t)\over\rho(r,t)}=(-t)^{-0.352}\,\left[{\delta D(\eta)\over
D(\eta)}\right]\,Y_{2m}, \\
&&{\delta {\bf v}(\br,t)\over v(r,t)}=(-t)^{-0.352}\,
\left[{\delta V_r(\eta)\over V(\eta)}\,{\hat r}+{\delta V_\perp(\eta)
\over V(\eta)}\hat\nabla_\perp\right]\,Y_{2m},
\ea
where $\rho(r,t)$ and $v(r,t)$ specify the unperturbed spherical flow,
and $\delta V_r(\eta)=\delta U'(\eta)$, $\delta V_\perp(\eta)=\delta
U(\eta)/\eta$. Figure 4 illustrates the growth of the density perturbation
as the collapse proceeds. The fractional perturbation grows as $(-t)^{-0.352}
\propto\rho_c(t)^{0.176}$, where $\rho_c(t)=\rho(0,t)$ is the central 
density of the cloud. The growing bar-mode
corresponds to the deformation of the collapsing cloud toward an ellipsoidal
shape. Depending on the initial perturbations, the deformed cloud may take the
form of an oblate disk or a prolate bar (see also Hanawa \& Matsumoto 1999). 

As $\gamma$ increases, the mode tends to be stablized by the effect of 
pressure. Figure 5 depicts the variation of $s=s_0$ 
for the lowest-order bar-mode
($l=2$) as a function of $\gamma$. We find that $s$ increases with
increasing $\gamma$, and the mode is unstable (with
negative $s$) only for $\gamma\le 1.09$.\footnote{Similar result
is also obtained by Hanawa \& Matsumoto (2000a) in a different
analysis. The author thanks the referee, T. Hanawa, for 
pointing out this paper.}
Figure 6 gives a few examples of the
mode eigenfunctions for several different values of $\gamma$. 

\subsubsection{Stable Modes}

We have searched numerically for other unstable modes [with Re$(s)<0$]
for $1\le \gamma\le 4/3$ and $l=1,2,3,\cdots$. Our search covers
the domain $-5\lo {\rm Im}(s)\lo 5$.
However, except for those discussed in \S 3.4.1, all modes we have found are
stable [with Re$(s)>0$]. As an example, the dashed curve in Fig.~6 shows
the eigenfunction of a high-order $l=2$ mode (for $\gamma=1$), with
$s_1=0.23+0.26i$. Note that as $\gamma$ increases, the ordering of the 
modes can change. This is seen from Figure 5: For $\gamma\lo 1.11$ we have
$s_0<$Re$(s_1)$, but for $\gamma\go 1.11$ we find $s_0>$Re$(s_1)$. 
We have not explored the spectrum of high-order 
modes in detail, since these modes are all stable. Moreover, 
as the fractional perturbations associated with the stable
modes diverge in the $\eta\rightarrow\infty$ limit (see eq.~[\ref{deldd}]),
these modes are only formally well-defined, but are of no physical 
importance.


\section{Perturbations of ``Inside-Out'' Collapse of Isothermal Cloud}

In this section we present our perturbation analysis of
Shu's expansion-wave solution which describes the ``inside-out'' collapse of
a isothermal gas cloud. The equation of state is $p=K\rho=\rho a^2$,
where $a$ is the sound speed.

\subsection{Spherical Inside-Out Collapse: Shu's 
Expansion-Wave Solution}

The expansion-wave solution describes the post-collapse ($t>0$) evolution
of the flow. The similarity variable is defined as
\be
\eta={r\over at}.
\ee
The flow variables can be written in self-similar forms as in equations
(\ref{rhot})-(\ref{ut}), except that in $\rho_t,\,v_t,\,p_t,\,\cdots$
we have to replace $(-t)$ by $t$ and set $\gamma=1$, i.e., 
\be
\rho_t={1\over 4\pi Gt^2},\quad
v_t=a,\quad
p_t={a^2\over Gt^2},\quad
\psi_t= a^2,\quad
m_t={a^3t\over G},\qquad
u_t=a^2t.
\ee
(Note that to follow Shu's convention, we have included the factor 
$4\pi$ in $\rho_t$.) In terms of the similarity variables,
the continuity equation, Euler equation, and Poisson equation are
\ba
&&(V-\eta)D'+DV'+2D\left({V\over\eta}-1\right)=0,\\
&&{D'\over D}+(V-\eta)V'+{M\over\eta^2}=0,\\
&&M'=\eta^2 D.
\ea
These equations can be rearranged into the standard form as given by Shu: 
\ba
&&\left[(V-\eta)^2-1\right]{D'\over D}
=(\eta-V)\left[D-2\left(1-{V\over\eta}\right)\right],\\
&&\left[(V-\eta)^2-1\right]V'
=(\eta-V)\left[D(\eta-V)-{2\over\eta}\right],
\ea
and $M=\eta^2(\eta-V)D$.

Some properties of the expansion-wave solution are as follows.
For $\eta>1$, the solution describes a static isothermal sphere, with
$V(\eta)=0$ and $D(\eta)={2/\eta^2}$. The surface $\eta=1$ is the rarefaction
(expansion) wave front. For $\eta\rightarrow 0$, the solution describes 
a free-fall, with  
$M\rightarrow M_0=0.975$,
$V\rightarrow -\left({2M_0/\eta}\right)^{1/2}$, and
$D\rightarrow \left({M_0/2\eta^3}\right)^{1/2}$.
While $D$ and $V$ are continuous at $\eta=1$, $D'$ and $V'$ are not:
\be
V'(1+)=0,~~D'(1+)=-4;\qquad
V'(1-)=1,~~D'(1-)=-2.
\label{discont}\ee
(The notation $\eta=1+$ means that $\eta\rightarrow 1$ from above, and
$\eta=1-$ means $\eta\rightarrow 1$ from below.)

\subsection{Perturbation Equations}

As in equations (\ref{delrho})-(\ref{delpsi0}), we consider perturbations
of the form
\ba
&&\delta\rho(\br,t)=t^s\rho_t\,\delta\!D(\eta)Y_{lm},\label{delrho1}\\
&&\delta u(\br,t)=t^s u_t\,\delta U(\eta)Y_{lm},\label{delu1}\\
&&\delta \psi(\br,t)=t^s\psi_t\,\delta \Psi(\eta)Y_{lm}.\label{delpsi1}
\ea
Since $\delta\rho(\br,t)/\rho(r,t)
=t^s\,\left[\delta D(\eta)/D(\eta)\right]Y_{lm}$ (and similarly for other 
variables), the power-law index $s$ specifies the evolution of 
the perturbation relative to the background: The flow is unstable if
Re$(s)>0$ and stable if Re$(s)<0$.
Substituting (\ref{delrho1})-(\ref{delpsi1}) into 
the perturbation equations (\ref{pert1})-(\ref{pert3}), we obtain
\ba
&&\hskip -0.7truecm
(V-\eta)\delta D'+D\delta U''+\left(-2+s+V'+{2\over\eta}V\right)\delta D
+\left(D'+{2\over\eta}D\right)\delta U'-{l(l+1)\over\eta^2}D\delta U=0,
\label{pp1}\\
&&\hskip -0.7truecm
(V-\eta)\delta U'+{\delta D\over D}+(1+s)\delta U+\delta\Psi=0,\label{p2}\\
&&\hskip -0.7truecm
\delta\Psi''+{2\over\eta}\delta\Psi'-{l(l+1)\over\eta^2}\delta\Psi
-\delta D=0.\label{p3}
\ea
We can use equation (\ref{p2}) to eliminate $\delta U''$ in equation 
(\ref{pp1}) and obtain 
\ba
&&\left[\left(V-\eta\right)^2-1\right]{\delta D'\over D}-(2V'+s)\delta U'
+\left[\Bigl(-2+s+V'+{2\over\eta}V\Bigr)(V-\eta)
+{D'\over D}\right]{\delta D\over D}\nonumber\\
&&\qquad\qquad\qquad-{l(l+1)\over\eta^2}(V-\eta)\delta U-\delta\Psi'=0.
\label{p1}\ea
Thus, the expansion-wave front ($\eta=1$) is a singular point of the 
perturbation equation. 
Also, equation (\ref{p3}) can be written in the integral form:
\be
\delta\Psi(\eta)=-\eta^l P(\eta)-{Q(\eta)\over\eta^{l+1}},
\label{delpsi2}\ee
where
\ba
P(\eta) &=& {1\over 2l+1}\int_\eta^\infty\!\!\eta^{\prime~1-l}
\delta D(\eta')\,d\eta',\qquad
P'=-{1\over 2l+1}\eta^{1-l}\delta D,\label{pprime}\\
Q(\eta) &=& {1\over 2l+1}\int_0^\eta\!\eta^{\prime~l+2}
\delta D(\eta')\,d\eta',\qquad
Q'={1\over 2l+1}\eta^{l+2}\delta D.\label{qprime}
\ea

\subsection{Series Solution for $\eta>1$}

For $\eta>1$, we have $V=0$ and $D=2/\eta^2$, the perturbation
equations can be solved in Frobenius series. We consider
the solution which satisfies $\delta D/D\rightarrow 0$, 
$\delta U\rightarrow 0$, and $\delta \Psi\propto\eta^{-l-1}\rightarrow 0$ 
for $\eta\rightarrow\infty$ (i.e., $\delta\Psi$ is given by the 
decreasing solution of the Laplace equation);\footnote{The fourth
order system of differential equations allows for four independent solutions,
but this solution (which must exist for any physical situation) alone  
is adequate for our stability analysis (\S 4.4).} The last
condition implies that $Q$ approaches a constant as $\eta\rightarrow\infty$.  
Thus we can write
\be
Q(\eta)=\sum_{n=0}^{\infty}q_{2n}\eta^{-2n}.
\ee
Equation (\ref{qprime}) then gives
\be
\delta D(\eta)=\sum_{n=0}^{\infty}d_{2n}\eta^{-2n-l-3},\qquad
d_{2n}=-2n(2l+1)\,q_{2n}.
\label{deld4}\ee
Using equations (\ref{delpsi2}), (\ref{pprime}) and requiring 
$P\rightarrow 0$ as $\eta\rightarrow \infty$, we have
\be
\delta\Psi(\eta)=\sum_{n=0}^{\infty}\psi_{2n}\eta^{-2n-l-1},\qquad
\psi_{2n}=-{2l+1\over 2n+2l+1}\,q_{2n}.
\label{delpsi3}\ee
Substituting (\ref{deld4}) and (\ref{delpsi3}) into (\ref{p2}) yields
\be
\delta U(\eta)=\sum_{n=0}^{\infty} u_{2n}\eta^{-2n-l-1},\qquad
u_{2n}={(2l+1)(2n^2+2ln+n+1)\over (2n+2l+1)(2n+l+2+s)}q_{2n}.
\ee
Finally, using equation (\ref{p1}), we obtain the recurrence relation:
\ba
&&(2n+3+l+s)d_{2n+2}=(2n+l+1)d_{2n}+2\left[
s(2n+l+1)+l(l+1)\right]u_{2n}\nonumber\\
&&\quad\qquad\qquad\qquad\qquad\qquad+2(2n+l+1)\psi_{2n}\qquad\qquad
\qquad\qquad\qquad
(n=0,1,2,\cdots)
\label{recur}
\ea
With this recurrence relation, the complete solution for $\eta>1$ 
can be obtained. Note that for $\eta\rightarrow\infty$,
the asymptotic scalings of the perturbations are
\be
\delta\Psi\rightarrow {q_0\over \eta^{l+1}},\qquad
\delta U\rightarrow {q_0\over (l+2+s)\eta^{l+1}},\qquad
\delta D\propto {1\over \eta^{l+5}}.
\ee

\subsection{Instability}

Here we use the series solution of \S 4.3 and the boundary condition
at the expansion-wave front ($\eta=1$) to show that Shu's solution is 
unstable. As equation (\ref{p1}) indicates, the expansion-wave front
is a singular point of the perturbation equation. A natural (necessary) 
boundary condition at $\eta=1$ is that the perturbation is finite (although
$\delta D$ and $\delta V_r=\delta U'$ can be discontinuous across
$\eta=1$; see below). 

We can examine the behavior of the perturbation at $\eta\rightarrow 1+$
using the series solution of \S 4.3. 
From the recurrence relation (\ref{recur}) we find, for $n\rightarrow\infty$,
\ba
&&{d_{2n+2}\over d_{2n}}\rightarrow 1-{1+s\over n},\\
&&{u_{2n+2}\over u_{2n}}\rightarrow 1-{2+s\over n},\\
&&{\psi_{2n+2}\over \psi_{2n}}\rightarrow 1-{3+s\over n}.
\ea
Thus in order for $\delta D$ to be finite at $\eta\rightarrow 1+$, 
we require ${\rm Re}(s)>0$ (e.g., Mathews \& Walker 1970). One can similarly
show that in order for $\delta V_r=\delta U'$ to be finite at 
$\eta\rightarrow 1+$, we require ${\rm Re}(s)>0$. 
Thus, any perturbations which are well-behaved at the expansion-wave
front must be globally unstable. 

A possible caveat in the analysis given above is that in the presence
of flow perturbations, the rarefaction front is also perturbed,
and $\delta D(\eta\rightarrow 1+)$ does not give the density 
perturbation at the perturbed expansion-wave front; one might therefore be
concerned that the divergence of $\delta D(\eta\rightarrow 1+)$ is a result 
of an improper definition of $\delta D$. To address this problem, 
we define a stretched radial coordinate via
\be 
\xi(\eta)=\eta \left(1+\Delta Y_{lm} t^s\right),
\ee
where $\Delta$ is a constant (to be determined). The perturbed rarefaction 
front is located at $\xi(\eta=1)$. 
Since $D(\eta)+\delta D(\eta)Y_{lm} t^s=D[\xi(\eta)]+\delta D[\xi(\eta)] 
Y_{lm}t^s$, we have
\be
\delta D[\xi(\eta)]=\delta D(\eta)-\eta D'(\eta)\Delta.
\label{deldxi}\ee
Similarly, $\delta U'[\xi(\eta)]=\delta U'(\eta)-\eta V'(\eta)\Delta$.
Since $\delta D[\xi(\eta)]$ and $\delta U'[\xi(\eta)]$
must be continuous across the rarefaction front, and since $D'$ and
$V'$ are discontinuous at $\eta=1$ (see eq.~[\ref{discont}]), we infer that 
$\delta D(\eta)$ and $\delta V_r(\eta)$ are discontinuous at $\eta=1$. 
Evaluating equation (\ref{p1}) at $\eta=1+$ and $\eta=1-$, 
we find
\be
\Delta=-{1\over s+1}\delta U'(1+).
\ee
Using equation (\ref{deldxi}) we obtain 
\be
\delta D[\xi(\eta=1+)]=\delta D(\eta=1+)-{4\over (s+1)}\delta U'(\eta=1+).
\ee
Using the series solution of \S 4.3, we can easily show that 
$\delta D[\xi(\eta=1+)]$ diverges unless Re$(s)>0$.

Another concern one might have is that the divergence of $\delta D$ and $\delta
V_r$ at $\eta=1+$ discussed above simply indicates that the series
expansion breaks down at $\eta=1$ rather than the actual divergence of
the function $\delta D$ and $\delta V_r$. To address this issue,
we show in Figure 7 several examples of the absolute value of
the density perturbation $\delta D$ at small $(\eta-1)$ for several 
different values of $s$. The function $\delta D$ is calculated using 
the series expansion given in \S 4.3 (normalized by setting $q_0=1$). 
We see that, in accordance with our discussion above, when 
Re$(s)<0$, the density perturbation $\delta D$ diverges as $\eta\rightarrow
1+$. Indeed, an analysis of the perturbation equations near 
$\eta=1+$ shows that for $0<x\equiv\eta-1\ll 1$ the perturbations have the
following behavior: 
\ba
&&\delta D=C_0x^s\left[1+{\cal O}(x)\right]+C_1\left[1+{\cal O}(x)\right],
\label{c0}\\
&&\delta U={C_0\over 2(s+1)}\,x^{s+1}\left[1+{\cal O}(x)\right]+C_2
\left[1+{\cal O}(x)\right],\\
&&\delta\Psi={C_0\over (s+1)(s+2)}\,x^{s+2}\left[1+{\cal O}(x)\right]+C_3
\left[1+{\cal O}(x)\right],
\label{c2}\ea
where $C_0,\,C_1,\,C_2,\,C_3$ are constants. This clearly shows
that $\delta D(\eta=1+)$ diverges for Re$(s)<0$ --- We could have
deduced this result simply by examing the perturbation equations 
near $\eta=1+$, except that without the series solution discussed in
\S 4.3 we would not know whether $C_0=0$ is a possibility.
The numerical results (based on the series solution) depicted in Figure 7
agree with (\ref{c0})-(\ref{c2}) and $C_0\ne 0$, i.e., 
the boundary condition at $\eta\rightarrow\infty$ requires
$C_0\ne 0$. It is this global consideration of the perturbations
at $\eta\rightarrow\infty$ and at $\eta\rightarrow 1+$ that forces us to
conclude that the expansion-wave solution is unstable to 
perturbations of all $l$'s. 

Note that our analysis above indicates Re$(s)>0$, but we have not 
solved for $s$. (The actual values of $s$ depend on the flow at $\eta<1$
and the boundary conditions at $\eta\rightarrow 0$.) 
Thus the growth rates of the instabilities are unknown at present.

\section{Discussion}

Early studies by Hunter (1962) and by Lin, Mestel \& Shu (1965)
demonstrated that uniform, pressure-free gas clouds undergoing
gravitational collapse are unstable toward fragmentation and 
shape deformation, with perturbations growing asymptotically as 
$\delta\rho(\br,t)/\rho(t)\propto (t_0-t)^{-1}\propto\rho(t)^{1/2}$ 
in the linear regime, where $t_0$ denotes the epoch of complete collapse, 
and $\rho(t)$ is the unperturbed uniform density. However, the presence
of even a small initial central concentration and pressure forces significantly
alters the evolution of the cloud. If the gas pressure is simply 
related to the density by a power-law, $p=K\rho^\gamma$ (polytropic
equation of state), the flow asymptotically approaches the
similarity solutions found by Larson (1969), Penston (1969) (for isothermal
gas $\gamma=1$), by Goldreich \& Weber (1980) (for $\gamma=4/3$), and by
Yahil (1983) (for general $\gamma$). Since the local Jeans length is of the
same order as the length scale at which the flow varies, a global analysis is 
needed to determine the stability properties of the collapsing cloud. 
The result (\S 3) presented in this paper (see also Hanawa \& Matsumoto 1999)
shows that for sufficiently soft equation of state ($\gamma\le 1.09$), the
Larson-Penston-Yahil similarity flow is unstable against bar-mode
perturbations, such that $\delta\rho(\br,t)/\rho(r,t)\propto
(t_0-t)^sY_{2m}(\theta,\phi)$ with 
$s<0$ ($s=-0.352$ for $\gamma=1$ and $s$ increases to zero as $\gamma$
increases to $1.09$, see Fig.~5), where $t_0$ denotes the epoch of core
formation. Since the central density increases as $\rho_c(t)\propto
(t_0-t)^{-2}$, the growth of perturbation, $\delta\rho(\br,t)/\rho(r,t)\propto 
\rho_c(t)^{-s/2}$, is slow (e.g., for isothermal collapse, 
$\delta\rho/\rho$ increases by a factor of $1.5$ when $\rho_c$ increases by
a factor of $10$). Such a slow growth (compared with the 
$\delta\rho/\rho\propto\rho^{1/2}$ behavior for the collapse of uniform,
pressure-less gas) is a result of the stablizing influence of pressure, 
despite the large Mach number (about $3$) achieved in the outer region of 
the cloud. 

Our stability analysis applies to the pre-collapse stage
(prior to core formation) of the Larson-Penston-Yahil solutions. 
After the central core forms, the outer core and envelope accrete 
onto it (see Fig.~1). The gas approaches free-fall as $r\rightarrow 0$, and the
Mach number becomes much greater than unity. In this (accretion) stage,
nonradial perturbations (of all scales) grow kinematically 
as $\delta\rho/\rho\propto r^{-1/2}\propto \rho^{1/3}$, where $r(t)$ is the
radius of a fluid element and $\rho(t)\propto r^{-3/2}$ its comoving density  
(Lai \& Goldreich 2000). Although the fluid element is free-falling, 
the perturbation grows more slowly compared with the case of uniform
pressure-less collapse because the steep velocity gradients provide
a stablizing influence on the flow. 

The global bar-mode instability for isothermal collapse may have 
important implications for star formation, particularly in connection with 
the formation of binary (and multiple) stars
(see also Hanawa \& Matsumoto 1999; Matsumoto \& Hanawa 1999).  
Fragmentation is unlikely to occur in a globally spherical collapse because  
small condensations do not contract fast enough to separate out from the
converging bulk flow. Angular momentum (or magnetic field) can obviously
make the cloud nonspherical, and thus facilitate fragmentation
(e.g., Burkert \& Bodenheimer 1996; Burkert, Bate \& Bodenheimer 1997; 
Truelove et al.~1997,1998; Boss 1998). Observations suggest that
many of the molecular cloud cores (with mass of order a few $M_\odot$
and size $0.1$~pc) have elongated shapes (Myers et al.~1991)
and slow rotation rates (with the ratio of rotational to gravitational energies
of order $0.02$; Goodman et al.~1993), implying that rotation is probably 
not a crucial factor in driving fragmentation on scales greater than 
$200$~AU. Our result on the growth of bar-mode perturbations 
($\delta\rho\propto Y_{2m}$) indicates that, even without net angular momentum,
the collapsing cloud tends to deform into an ellipsoidal shape (oblate
disk or prolate bar, depending on which $m$-mode perturbation is 
dominant initially). Fragmentation is more likely to occur for such deformed
configurations (e.g., Bonnel 1999; Matsumoto \& Hanawa 1999). 

In the context of core-collapse supernovae, our result shows that 
the homologous inner core and the supersonic outer core are 
globally stable against nonradial perturbations prior to core bounce at 
nuclear density and the formation of the proto-neutron star.
However, during the subsequent accretion of the outer core (involving $15\%$ of
the core mass) and envelope onto the proto-neutron star, nonspherical
perturbations can grow according to $\delta\rho/\rho\propto r^{-1/2}$ or even
$\delta\rho/\rho \propto r^{-1}$ (Lai \& Goldreich 2000). 
The asymmetric density perturbations seeded in the presupernova star,
especially those in the outer region of the iron core, are therefore amplified
during collapse. The enhanced asymmetric density perturbation may lead to
asymmetric shock propagation and breakout, which then give rise to asymmetry 
in the explosion and a kick velocity to the neutron star (Goldreich et
al.~1996; Burrows \& Hayes 1996).  

Our stability analysis (\S 4) shows that Shu's expansion-wave solution
is globally unstable to perturbations of all $l$'s,
although the growth rates are unknown at present.
The implication of this result is not entirely clear. 
It is well-known that a static singular isothermal sphere
is highly unstable to radial perturbations 
(A truncated Bonner-Ebert isothermal sphere is
unstable when the range of density from the center to the surface is greater
than $14.04$; see Bonner 1956, Hunter 1977). 
Earlier one-dimensional numerical simulations have already shown 
that a collapsing isothermal cloud does not approach the
expansion-wave solution (Hunter 1977; Foster \& Chevalier 1993).
Our stability analysis corroborates this result, and indicates that 
the expansion-wave solution cannot be realised in a pure hydrodynamical 
situation. 

Magnetic fields play an important role in the current paradigm for 
forming low-mass stars (e.g., Shu, Adams \& Lizano 1987;
Shu et al.~1999; Mouschovias \& Ciolek 1999). 
Ambipolar diffusion of magnetic fields drives the quasi-static contraction of
the molecular cloud core with growing central concentration such that the core
asymptotically approaches the state of a singular isothermal sphere.
When the flux-to-mass ratio drops below certain critical value, a runaway
``inside-out'' collapse ensues, and it is thought that this collapse 
is well described by the expansion-wave solution (Shu et al.~1999).
In reality, there is probably no sharp distinction between the quasi-static
contraction and dynamical collapse (e.g., Safier, McKee \& Stahler 1997; Li
1998), and a real singular isothermal sphere can never be reached. 
Our global stability analysis of the expansion-wave solution (\S 4) does not 
depend on the mathematical singularity of the solution at $r=0$,
but depends on the existence of a well-defined rarefaction front
and a static isothermal density profile outside the front in the 
solution. It in not clear whether our idealized hydrodynamical stability
analysis can be applied to more realistic situations with 
(even sub-dominant) magnetic fields (see Galli \& Shu 1993a,b and Li \& Shu 
1997 for the effects of magnetic field on self-similar ``inside-out'' 
collapse). 

\acknowledgments
This work was started in 1995 when I was a postdoc in theoretical 
astrophysics at Caltech (support from a Richard C. Tolman fellowship is
gratefully acknowledged). I thank Peter Goldreich for initially suggesting 
this problem in the context of core-collapse supernovae and for 
many valuable discussions. I also thank Frank Shu and the
referee, T. Hanawa, for useful comments on this paper. 
This work is supported in part
by NASA grants NAG 5-8356 and NAG 5-8484, and by a research fellowship from the
Alfred P. Sloan foundation.   

\appendix

\section{Rotational Perturbations in Larson-Penston-Yahil Solutions}

A general velocity perturbation can be written as
\be
\delta\bv(\br,t)=\delta v_r(r,t)Y_{lm}\,\hat r+\delta v_\perp(r,t)
\hat\nabla_\perp Y_{lm}+\hat\nabla_\perp\times\left[\delta v_{\rm rot}
(r,t)Y_{lm}\,\hat r\right].
\ee
Using Euler equation, we obtain
\ba
&&{d\over dt}(r\delta v_{\rm rot})=0,\label{dvrot}\\
&&{d\over dt}\delta v_T=-{\partial v\over\partial r}\delta v_T,
\label{dvt}
\ea
where $\delta v_T\equiv\delta v_r-\partial (r\delta v_\perp)/\partial r$
(see Lai \& Goldreich 2000). The potential flow discussed in the main
text corresponds to $\delta u=r\delta v_\perp$, $\delta v_T=0$ and
$\delta v_{\rm rot}=0$.
Note that $\delta v_{\rm rot}$ is decoupled from the potential flow. 

Writing $\delta v_{\rm rot}$ in the self-similar form,
$\delta v_{\rm rot}(r,t)=v_t (-t)^s\delta V_{\rm rot}(\eta)$,
equation (\ref{dvrot}) becomes
\be
W(\eta\delta V_{\rm rot})'+(-s-3+2\gamma)(\eta\delta V_{\rm rot})=0,
\label{vrot0}\ee
where $W=V+(2-\gamma)\eta$. 
Since $V\propto \eta$ as $\eta\rightarrow 0$, it is most natural to require
$\delta V_{\rm rot}\propto\eta$ at $\eta\rightarrow 0$ (corresponding to
a uniform ``rotation''). Equation (\ref{vrot0}) then gives
$s=-1/3$, independent of $\gamma$. This is a growing mode
which describes the spin-up of a rotating cloud during gravitational collapse
(see Hanawa \& Nakayama 1997; Matsumoto \& Hanawa 1999). The ``angular
frequency'' increases as 
$
{\delta v_{\rm rot}/ r}\propto  (-t)^{-4/3}
\,{\delta V_{\rm rot}/\eta}$,
and the velocity perturbation increases as $\delta v_{\rm rot}/v\propto
(-t)^{-1/3}\propto\rho_c(t)^{1/6}$.

Similarly, writing $\delta v_T$ as
$\delta v_T(r,t)=v_t (-t)^s\delta V_T(\eta)$,
equation (\ref{dvt}) becomes
\be
W\delta V_T'+(V'+\gamma-1-s)\delta V_T=0.
\label{delvt}\ee
For $\eta\rightarrow 0$, we have $\delta V_r\propto \eta^{l-1}$,
$\delta V_\perp\propto\eta^{l-1}$, but $\delta V_T\propto \eta^{l+1}$.
Equation (\ref{delvt}) then gives
$s=(4/3-\gamma)l-1/3$. This is the growing ``votex'' mode discussed
by Hanawa \& Matsumoto (2000b).

\clearpage

\begin{deluxetable}{rrrr}
\tablecolumns{4}
\tablewidth{0pc}
\tablecaption{Parameters for Pre-collapse Larson-Penston-Yahil Solutions}
\tablehead{
\colhead{$\gamma$} & \colhead{$D_0$} & \colhead{$\eta_s$} &
\colhead{${\cal M}_\infty$} 
}
\startdata
1.00 & 0.13256 & 2.34113 & 3.271\\
1.05 & 0.18299 & 2.33723 & 2.802\\
1.10 & 0.25908 & 2.33252 & 2.506\\
1.15 & 0.37769 & 2.32991 & 2.338\\
1.20 & 0.57228 & 2.33277 & 2.290\\
1.25 & 0.92455 & 2.34606 & 2.407\\
1.30 & 1.75375 & 2.38512 & 2.944\\
1.31 & 2.10358 & 2.40203 & 3.216\\
1.32 & 2.66174 & 2.42909 & 3.695\\
1.33 & 3.99500 & 2.49457 & 5.113\\
\enddata
\tablecomments{$\gamma$ is the polytropic index, $D_0=D(\eta=0)$,
$\eta_s$ is the sonic point, where $W=A=(\gamma D^{\gamma-1})^{1/2}$,
and ${\cal M}_\infty=(|V|/A)_\infty$ is the Mach number at $\eta\rightarrow
\infty$.}
\end{deluxetable}

\clearpage

\begin{figure}
\begin{center}
\hbox{\epsfysize=5.0in
\epsffile{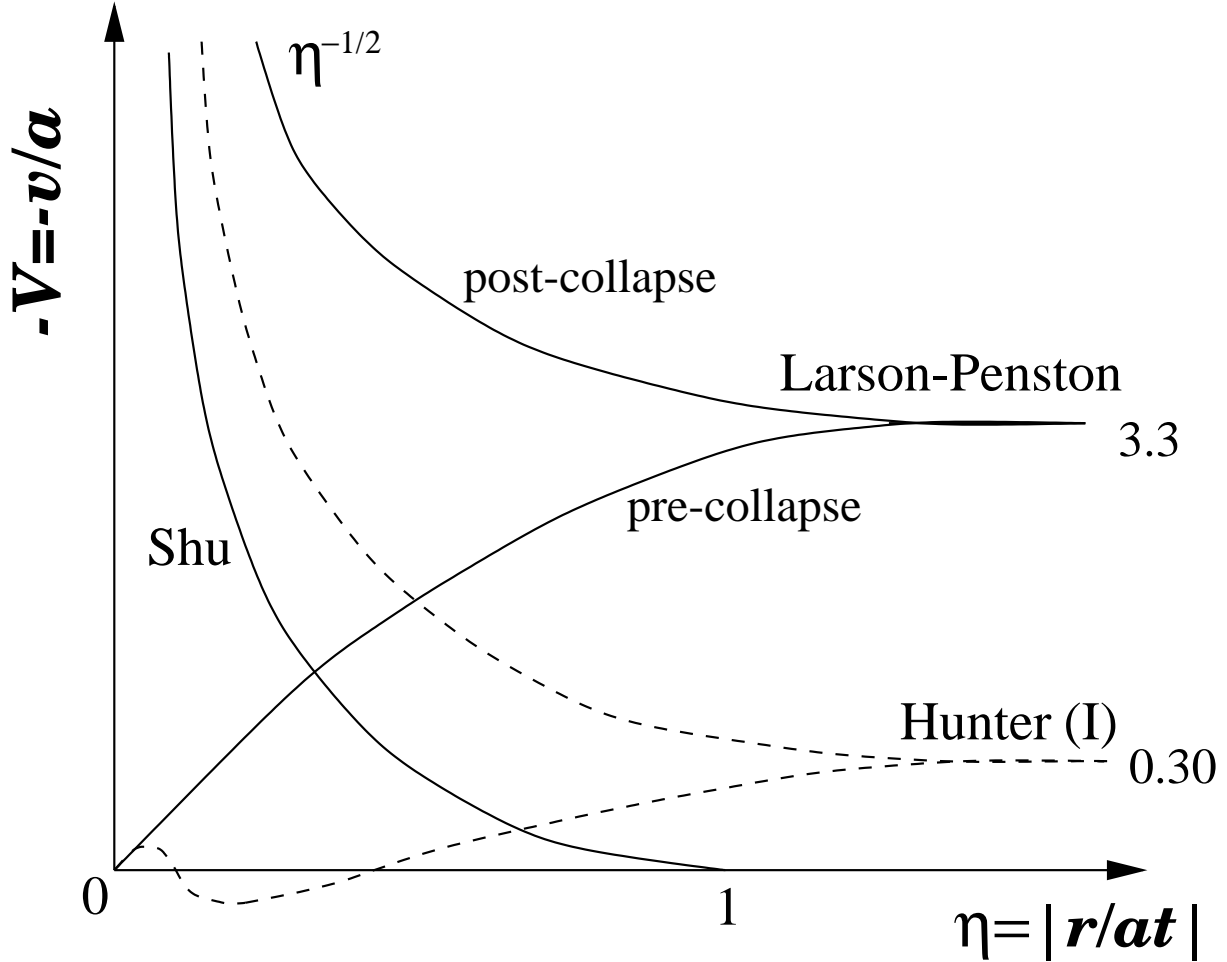}}
\caption{
A schematic diagram showing the properties of different 
self-similar solutions describing the collapse and accretion 
of isothermal gas clouds. The similarity variable is $\eta=r/(-at)$
for pre-collapse solutions ($t<0$) and $\eta=r/(at)$ for post-collapse 
(accretion) solutions ($t>0$), where $a$ is the sound speed, and $t=0$
corresponds to the epoch when the central core collapses to form a protostar. 
The vertical 
axis gives the dimensionless inflow flow velocity $-V=-v/a$. 
Shu's expansion-wave solution (the solid curve that terminates
at $\eta=1$, the rarefaction wave front) describes post-collapse 
accretion, while the other solutions have a pre-collapse phase
and a post-collapse phase which are connected at $\eta\rightarrow\infty$
(or $t=0$). All post-collapse flows approach free-fall
$V\propto\eta^{-1/2}$ as $\eta\rightarrow 0$. At $\eta\rightarrow\infty$, the
Larson-Penston solution has Mach number of $3.3$. The dashed curves
give an examples of the infinite (but discrete) number of type I 
solutions found by Hunter (1977). (Note that the pre-collapse
type I solutions contain regions with both positive and
negative $v$.)
 The expansion-wave solution is the limiting
case ($V\rightarrow 0$ at $\eta\rightarrow\infty$) of the post-collapse
Type I solutions. Note that all pre-collapse solutions have $V\rightarrow
-2\eta/3$ as $\eta\rightarrow 0$. 
}\end{center}
\end{figure}

\begin{figure}
\begin{center}
\hbox{\epsfysize=7.0in
\epsffile{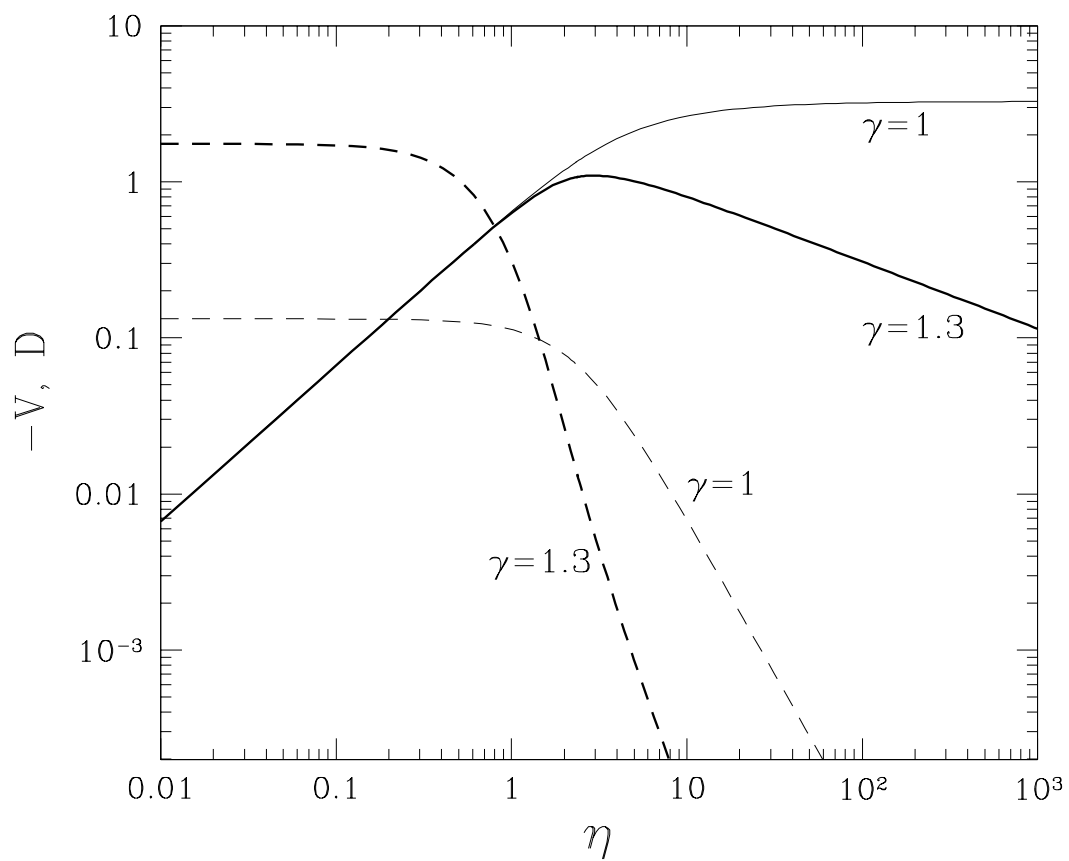}}
\caption{
The Larson-Penston-Yahil similarity solutions are shown for $\gamma=1$
and $\gamma=1.3$. The solid curves give the dimensionless flow velocity
$(-V)$, and the dashed curves give the density profile $D$. 
}\end{center}
\end{figure}

\begin{figure}
\begin{center}
\hbox{\epsfysize=7.0in
\epsffile{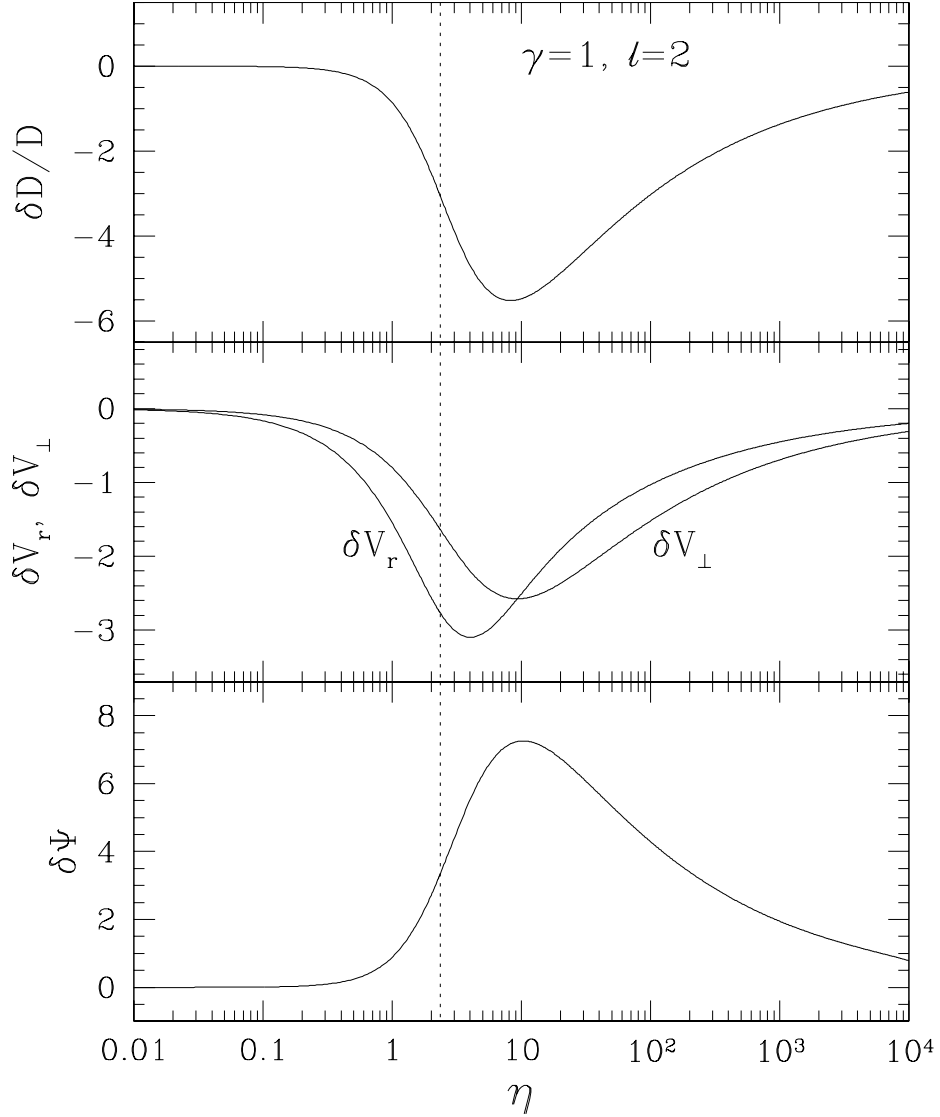}}
\caption{
Eigenfunctions of the lowest-order bar-mode ($l=2$) for $\gamma=1$
(isothermal collapse), with the eigenvalue
$s=-0.352$. The upper panel shows the fractional density 
perturbation $\delta D/D$, the middle panel shows the radial and
tangential velocity perturbations, and the lower panel shows the
potential perturbation. The dotted vertical line denotes
the transonic point $\eta_s=2.341$. The similarity variable is 
$\eta=r/(-at)$, where $a$ is the (isothermal) sound speed. 
}\end{center}
\end{figure}

\begin{figure}
\begin{center}
\hbox{\epsfysize=7.0in
\epsffile{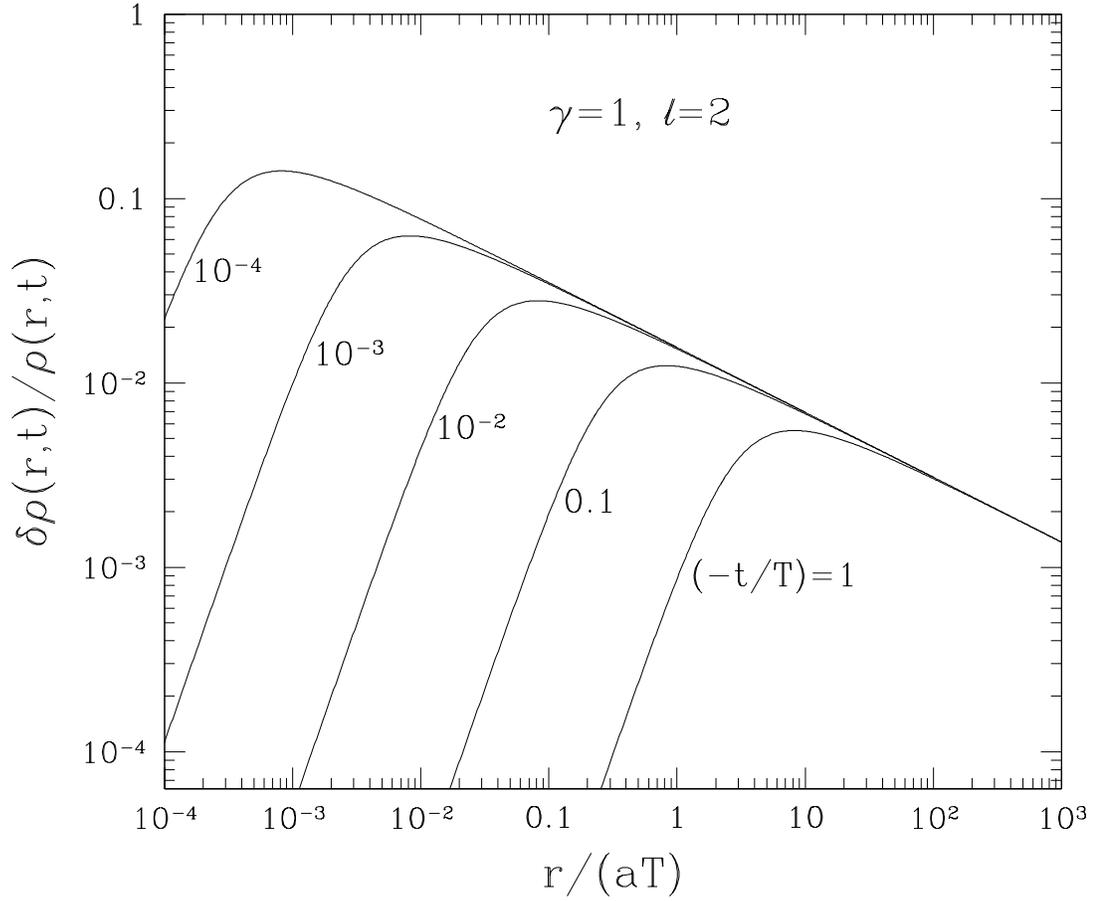}}
\caption{
Evolution of the $l=2$ density perturbation during the collapse of
an isothermal cloud ($\gamma=1$). The angular dependence, $Y_{2m}$,
has be suppressed. Note that $\delta\!\rho(r,t)/\rho(r,t)\propto (-t)^{-0.352}
\delta\!D(\eta)/D(\eta)$, with $\eta=r/(-at)$; $T$ is a fiducial 
time, and $a$ is the sound speed. The different curves correspond to
different times. The center of the cloud reaches singularity
as $t$ approaches zero. 
}\end{center}
\end{figure}

\begin{figure}
\begin{center}
\hbox{\epsfysize=7.0in
\epsffile{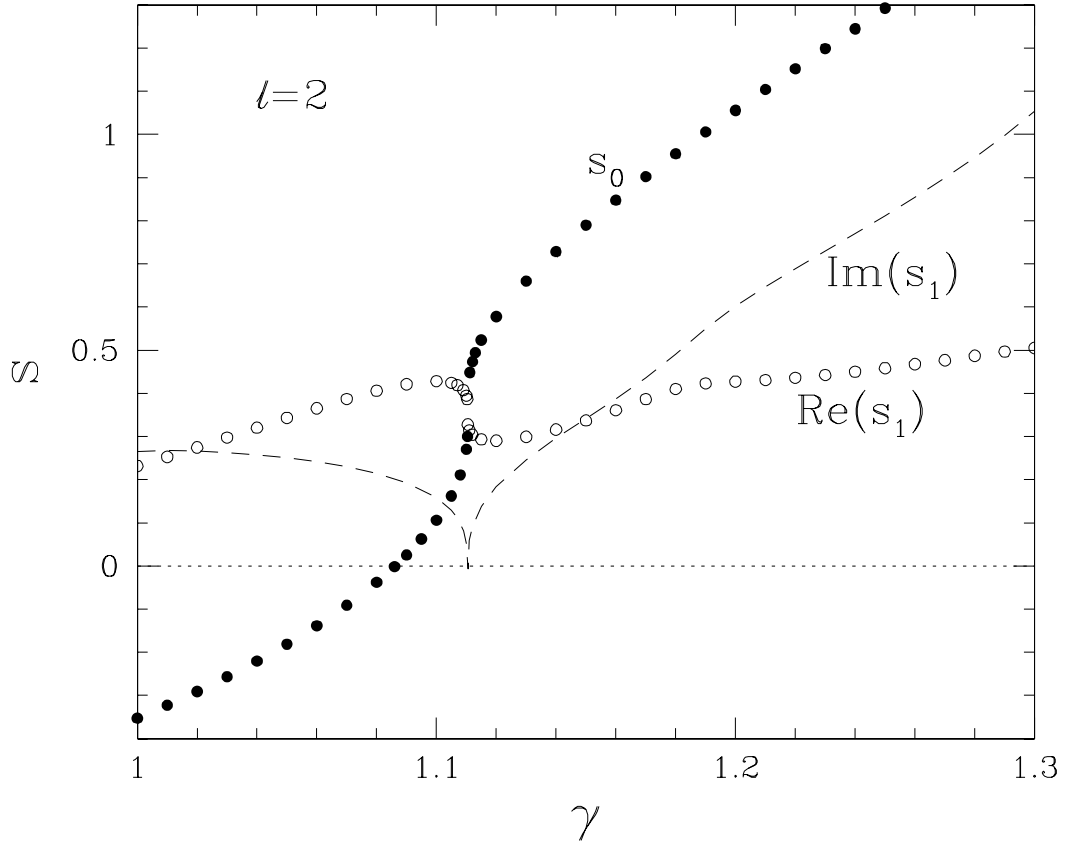}}
\caption{
The eigenvalue $s$ of the bar-mode ($l=2$) as a function of the polytropic
index $\gamma$. The filled circles correspond to the lowest-order 
mode (with no radial node in the eigenfunction), with $s=s_0$ real; The 
open circles correspond to Re$(s_1)$ of a higher-order mode,
with $s=s_1$ complex, the dashed curve gives Im$(s_1)$ of the same mode. 
Note that the lowest-order bar mode is globally unstable
for $\gamma\le 1.09$. The $s=s_1$ mode has one radial node for
$\gamma\le 1.15$, and it crosses the zero-node mode ($s=s_0$) 
at $\gamma\simeq 1.11$. For $\gamma\ge 1.11$, the $s=s_1$ mode
is the mode with the lowest Re$(s_1)$.
}\end{center}
\end{figure}

\begin{figure}
\begin{center}
\hbox{\epsfysize=7.0in
\epsffile{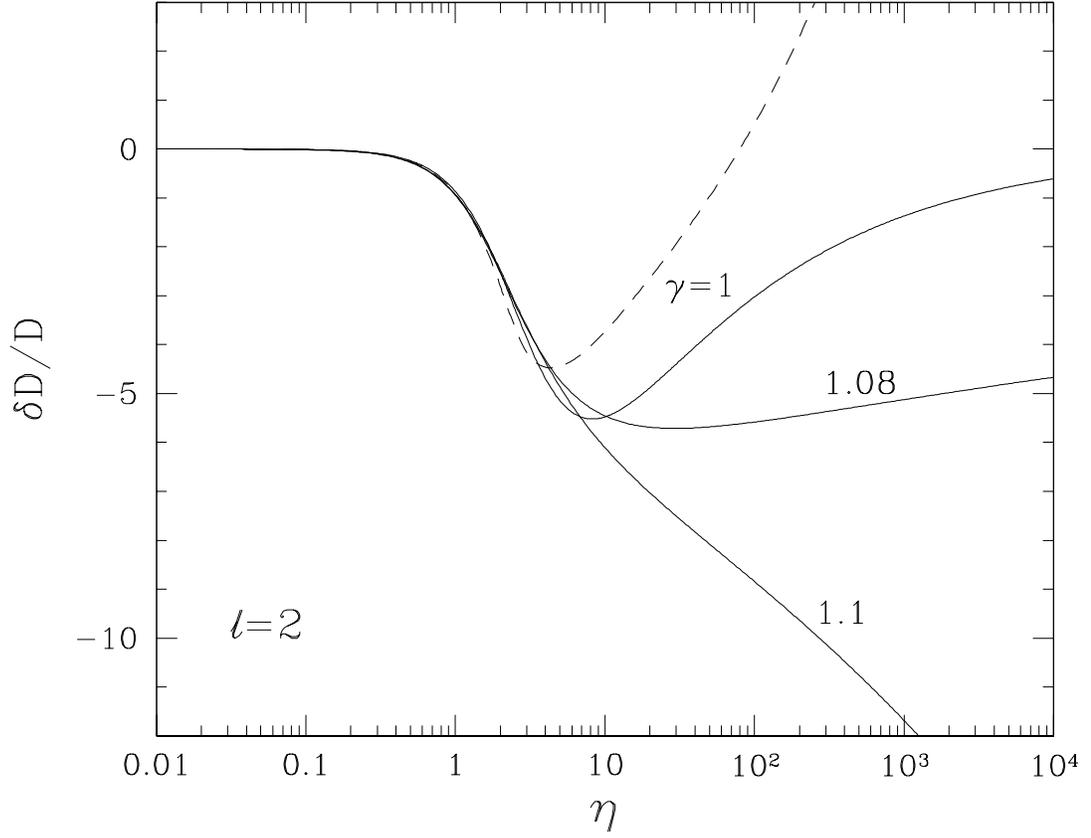}}
\caption{
The eigenfunctions $\delta D/D$ of several bar-modes ($l=2$)
for different $\gamma$. The solid curves correspond to the
lowest-order bar-mode for $\gamma=1$ (with $s=-0.352$, unstable), 
$\gamma=1.08$ (with $s=-0.038$, unstable), and $\gamma=1.1$ 
(with $s=0.106$, stable). The dashed curve gives
Re($\delta D/D$) for the $s=s_1=0.23+0.26i$ mode (see Fig.~5) with $\gamma=1$.
}\end{center}
\end{figure}


\begin{figure}
\begin{center}
\hbox{\epsfysize=7.0in
\epsffile{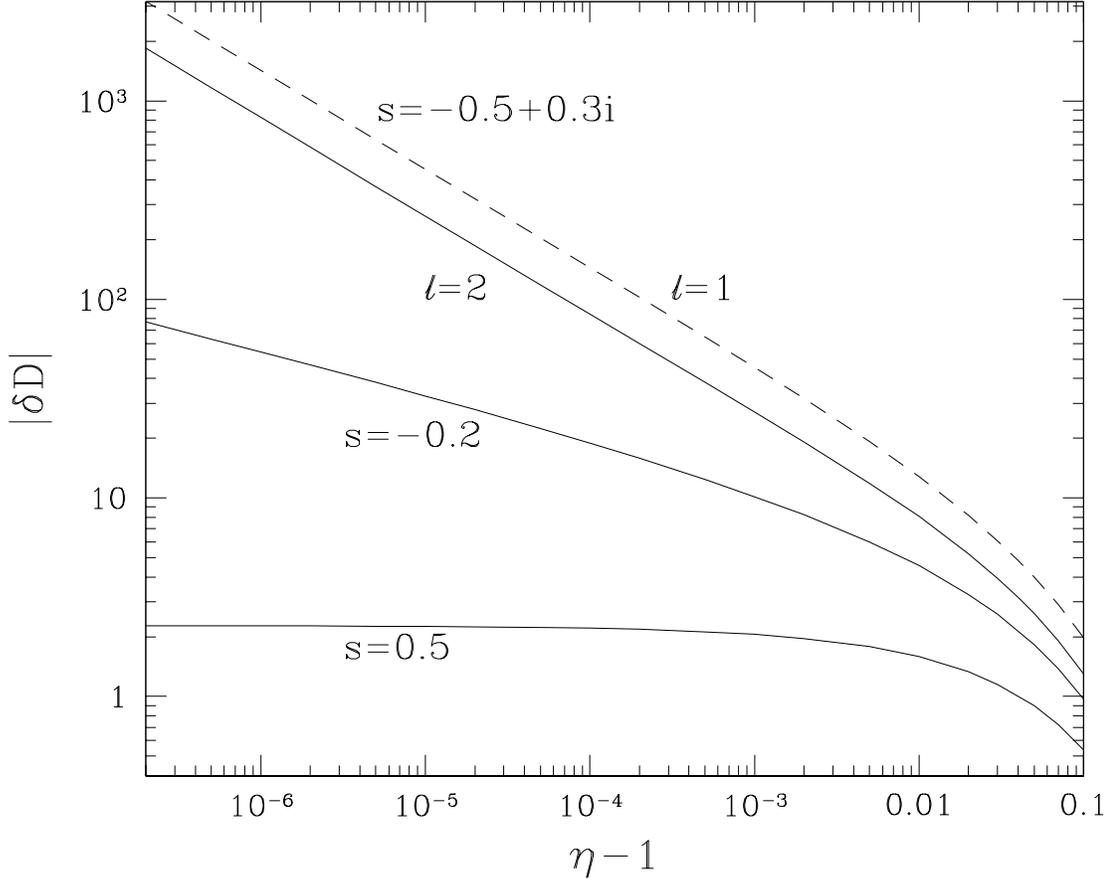}}
\caption{
Behavior of the absolute value of the density perturbation $\delta D$ 
just outside the expansion-wave front ($\eta=1$) in the expansion-wave 
solution. The solid curves are for $l=2$ and the dashed curve for $l=1$.
The values of $s$ are labeled for each curve. Note that when 
Re$(s)<0$, the perturbation $|\delta D| \rightarrow\infty$ as 
$\eta\rightarrow 1$.
}\end{center}
\end{figure}

\end{document}